\documentclass[epsf]{article}
\usepackage{graphicx}
\usepackage{amsmath}
\usepackage{epsf}

\input{tcilatex}

\begin{document}

\title{The influence of the oscillations of the chemical potential on the de Haas -
van Alphen effect in quasi-two-dimensional compounds \\
}
\author{P.Grigoriev \\
%EndAName
L.D.Landau Institute for Theoretical Physics \\
142432, Chernogolovka, Moscow region, Russia \\
e-mail: pashag@itp.ac.ru\\
Grenoble High Magnetic Field Laboratory \\
MPI-FKF and CNRS \\
BP 166, F-38042 Grenoble Cedex 09, France}
\maketitle

\begin{abstract}
The de Haas - van Alphen effect in quasi-two-dimensional metals is studied
at arbitrary parameters. The oscillations of the chemical potential may
substantially change the temperature dependence of harmonic amplitudes that
is usually used to determine the effective electron mass. Hence, the
processing of the experimental data using the standard Lifshitz-Kosevich
formula (that assumes the chemical potential to be constant) may lead to
substantial errors even in the limit of strong harmonic damping. This fact
may explain the difference between the effective electron masses, determined
from the de Haas - van Alphen effect and the cyclotron resonance
measurements. The oscillations of the chemical potential and the deviations
from the Lifshitz-Kosevich formula depend on the reservoir density of
states, that exists in organic metals due to open sheets of Fermi surface.
This dependence can be used to determine the density of electron states on
open sheets of Fermi surface. We present the analytical results of the
calculations of harmonic amplitudes in some limiting cases that show the
importance of the oscillations of the chemical potential. The algorithm of
the simple numerical calculation of the harmonic amplitudes at arbitrary
reservoir density of states, arbitrary warping, spin-splitting, temperature
and Dingle temperature is also described.
\end{abstract}

The quantum magnetization oscillations (or the de Haas - van Alphen (dHvA)
effect) was discovered long ago\cite{dHvA} and has been used a lot as a
powerful tool for studying the Fermi surfaces and single electron properties
in metals\cite{Sh}. In three dimensional (3D) metal the good quantitative
description of this effect is given by the Lifshitz-Kosevich(L-K) formula 
\cite{LK}. In two- or quasi-two-dimensional compounds the deviations from
the L-K formula are possible for three reasons: the harmonic damping in
two-dimensional (2D) case is different, the impurity scattering may not be
described by the usual Dingle law and the chemical potential becomes also an
oscillating function of the magnetic field. The first problem is important
only when the harmonic damping is weak and can be easily solved by using the
2D harmonic expansion\cite{Sh}. The second problem concerns with an accurate
calculation of the density of states (DoS) with electron-electron
interactions and the impurity scattering. The electron-electron interactions
are not very important when many Landau levels are occupied (we consider the
case when the Fermi energy $\epsilon _{F}$ is much greater than the Landau
level separation and temperature). The impurity scattering in 3D case adds
an imaginary part $i\Gamma (B)$ to the electron spectrum that means that an
electron may leave its quantum state with probability $w=\Gamma (B)/\pi
\hbar $ per second. If one assumes this width $\Gamma (B)$ of energy levels
to be independent of the magnetic field $B$ he gets the Dingle law of
harmonic damping\cite{Dingle} 
\begin{equation*}
A_{l}\sim \exp \left( -2\pi l\Gamma /\hbar \omega _{c}\right)
\end{equation*}
where $A_{l}$ is the amplitude of the harmonic number $l$ and $\omega
_{c}=eB/m^{\ast }c$\ is the cyclotron frequency. This Dingle law has been
proved by many experiments on 3D metals. In 2D case this law may be
incorrect and the problem of the DoS distribution in 2D metals has not been
solved yet, although many theoretical works have been devoted to this
subject (for example, \cite{TA},\cite{M},\cite{VM}). The problem is
complicated because even the exact calculation of the point-like impurity
scattering is not enough since the long-range impurities (and, probably, the
electron-electron interactions) are also important in 2D case\cite{XLi}. The
procedure of extracting the DoS distribution from the dHvA measurements has
been proposed recently\cite{we}. \ In the present paper we focus on the
third question, so, we assume the Dingle law to be valid and in this
approximation we consider the influence of the oscillations of the chemical
potential on the harmonic amplitudes of the dHvA oscillations. Since we
consider the quasi-2D case, the Dingle law is not a bad approximation. We
shall show that the oscillations of the chemical potential change
substantially the temperature and Dingle temperature dependence of the
harmonic amplitudes even in the limit of strong harmonic damping. Hence, the
estimate of the effective electron mass based on the L-K formula may lead to
the errors up to 30\%. This can be an explanation of the difference of the
effective electron masses obtained from the dHvA effect and cyclotron
resonance measurements (for example, in \cite{Wos} and \cite{CR}). This
problem was examined numerically by N. Harrison et al.\cite{Harr} at zero
warping $W$ of Fermi surface(FS). In this paper we derive the explicit
formulas that describe the quantum magnetization oscillations at arbitrary
parameters. The analytical study of this result is made in some limiting
cases. It shows the importance of the oscillations of chemical potential on
the harmonic amplitudes.

The energy spectrum of quasi-two-dimensional electron gas is 
\begin{equation}
E_{n,k_{z},\sigma }=\hbar \omega _{c}\,(n+\frac{1}{2})+\frac{W}{2}\cos
(k_{z}d)+\sigma \mu _{e}B  \label{Enk}
\end{equation}
where $W$ is the warping of quasi-cylindrical Fermi surface. The DoS
distribution with impurity scattering may be written as 
\begin{equation*}
\rho (E,B)=\rho _{0}(E,B)+\tilde{\rho}(E,B)
\end{equation*}
where the oscillating part of the DoS at $E\gg \hbar \omega _{c}$ is\cite
{Min} 
\begin{equation}
\tilde{\rho}(E,B)=\frac{4g}{\hbar \omega _{c}}\sum_{l=1}^{+\infty
}(-1)^{l}\cos \left( 2\pi l\frac{E}{\hbar \omega _{c}}\right) \,J_{0}\left(
\pi l\frac{W}{\hbar \omega _{c}}\right) \cos \left( 2\pi l\frac{\mu _{e}B}{%
\hbar \omega _{c}}\right) \,\exp \left( -\frac{2\pi l\,\Gamma }{\hbar \omega
_{c}}\right)  \label{rh}
\end{equation}
In this formula $g=B/\Phi_0$ is the Landau level(LL) degeneracy, the factor $%
\cos \left( 2\pi l \mu _{e}B/\hbar \omega _{c}\right)$ is due to
spin-splitting and the factor $J_{0}\left( \pi l W/\hbar \omega _{c}\right)$
comes from the finite warping $W$ of quasi-cylindrical Fermi surface. $%
J_0(x) $ is the Bessel function of zeroth order. The last factor in (\ref{rh}%
) is the usual Dingle factor.

The non-oscillating part of the DoS 
\begin{equation*}
\rho _{0}(E,B)=\frac{2g}{\hbar \omega _{c}}(1+n_{R}(E))
\end{equation*}
where $n_{R}(E)$ is the ratio of the reservoir density of states to the
average DoS on quasi-2D part of FS. The reservoir density of states exists
in quasi-2D organic metals due to open sheets of Fermi surface. These
quasi-one-dimensional states do not contribute to magnetization oscillations
since they form a continuous spectrum.

If the DoS is known one can calculate the thermodynamic potential 
\begin{equation}
\Omega (\mu ,B,T)=-T\int_{0}^{\infty }\rho (E,B)\ln \left[ 1+\exp \left( 
\frac{\mu -E}{T}\right) \right] \,dE=\Omega _{0}(\mu ,B,T)+\tilde{\Omega}%
(\mu ,B,T)  \label{om}
\end{equation}
where $\mu (B)$ is the chemical potential and the oscillating part of the
thermodynamic potential is\cite{Min} 
\begin{equation*}
\tilde{\Omega}=2gT\sum_{l=1}^{+\infty }\frac{(-1)^{l}}{l} \cos (2\pi l\frac{%
\mu }{\hbar \omega _{c}})\frac{ \lambda l}{\sinh (\lambda l)}J_{0}(\pi l%
\frac{W}{\hbar \omega _{c}})\cos (2\pi l\frac{\mu _{e}H}{\hbar \omega _{c}}%
)\,\exp \left( -\frac{2\pi l\,\Gamma }{\hbar \omega _{c}}\right)
\end{equation*}
where $\lambda \equiv 2\pi ^{2}T/\hbar \omega _{c}.$ The total particle
number is usually constant: 
\begin{equation*}
N=-\left( \frac{\partial \Omega (\mu ,B,T)}{\partial \mu }\right)
_{T,B}=\int_{0}^{\infty }\frac{\rho (E,B)}{1+\exp \left( \frac{E-\mu }{T}%
\right) }\,dE=const
\end{equation*}
This is an equation on the chemical potential as a function of magnetic
field. Separating the oscillating part of the DoS and substituting 
\begin{equation*}
N=\int_{0}^{\infty }\frac{\rho _{0}(E,B)}{1+\exp \left( \frac{E-\varepsilon
_{F}}{T}\right) }\,dE
\end{equation*}
($\varepsilon_{F}$ is the Fermi energy at zero magnetic field) we get 
\begin{equation}
\int_{0}^{\infty }\left( \frac{1}{1+\exp \left( \frac{E-\varepsilon _{F}}{T}%
\right) }-\frac{1}{1+\exp \left( \frac{E-\mu }{T}\right) }\right) \rho
_{0}(E,B)\,dE=\int_{0}^{\infty }\frac{\tilde{\rho}(E,B)}{1+\exp \left( \frac{%
E-\mu }{T}\right) }\,dE  \label{eq}
\end{equation}
Now we use the fact that the reservoir DoS $n_{R}(E)$\ does not change
appreciably on the scale of $T$ or $\vert\mu -\varepsilon _{F}\vert <\hbar
\omega _{c}/2$\ (this is true if many LLs are occupied because $n_{R}(E)$
changes substantially on the scale of Fermi energy). Then $n_{R}(E)\approx
n_{R}(\varepsilon _{F})=const\equiv n_{R}$ . The left hand side of equation (%
\ref{eq}) now simplifies and after substitution of (\ref{rh}) we get the
equation on the oscillating part $\tilde{\mu}(B)$ of the chemical potential 
\begin{equation}
\tilde{\mu}(B)\equiv \mu (B)-\varepsilon _{F}=\frac{\hbar \omega _{c}}{\pi
(1+n_{R}(\varepsilon _{F}))}\times  \label{muh}
\end{equation}
\begin{equation*}
\times \sum_{l=1}^{+\infty }\frac{(-1)^{l+1}}{l}\sin \left( \frac{2\pi
l\,(\varepsilon _{F}+\tilde{\mu}(B))}{\hbar \omega _{c}}\right) \frac{%
\lambda l}{\sinh (\lambda l)}\cos \left( 2\pi l\frac{\mu _{e}H}{\hbar \omega
_{c}}\right) \exp \left( -\frac{2\pi l\,\Gamma }{\hbar \omega _{c}}\right)
J_{0}\left( \pi l\frac{W}{\hbar \omega _{c}}\right)
\end{equation*}
This nonlinear equation can not be solved analytically without any
approximations but it determines the oscillations of the chemical potential
arbitrary parameters (only $\epsilon_F \gg T, \hbar\omega_c$ is assumed).

The magnetization oscillations at constant electron density $N=const$ 
\begin{equation*}
M=-\frac{d(\Omega +N\mu )}{dB}\mid _{N=const}=-\frac{\partial \,\Omega }{%
\partial B}\mid _{\mu ,N=const}-
\end{equation*}
\begin{equation*}
-\left( \frac{\partial \,\Omega }{\partial \mu }\mid _{N,B=const}+N\right) 
\frac{d\mu }{dB}\mid _{N=const}=-\frac{\partial \,\Omega }{\partial B}\mid
_{\mu ,N=const}
\end{equation*}
The oscillating part of the magnetization 
\begin{eqnarray}
\tilde{M}(B) &=&-\frac{\partial \,\tilde{\Omega}}{\partial B}\mid _{\mu
,N=const}=\frac{2g}{\pi B}\varepsilon _{F}\sum_{l=1}^{+\infty }\frac{%
(-1)^{l+1}}{l}\frac{\lambda l}{\sinh \lambda l}\cos (2\pi l\frac{\mu _{e}H}{%
\hbar \omega _{c}})\exp \left( -\frac{2\pi l\,\Gamma }{\hbar \omega _{c}}%
\right) \times  \label{M} \\
&&\times \left\{ \sin (2\pi l\frac{\mu (B)}{\hbar \omega _{c}})J_{0}(\pi l%
\frac{W}{\hbar \omega _{c}})+\frac{W}{2\mu }\cos (2\pi l\frac{\mu (B)}{\hbar
\omega _{c}})J_{1}(\pi l\frac{W}{\hbar \omega _{c}})\right\} \,  \notag
\end{eqnarray}
where $\mu (B)$ is given by equation (\ref{muh}) and contains the dependence
of magnetization on the reservoir DoS. The formulas (\ref{muh}) and (\ref{M}%
) describe the magnetization oscillations at arbitrary parameters. The only
approximation, used in these formulas is the Dingle law of harmonic damping.
In quasi-2D organic metals with warping $W>T_D$ the Dingle law is believed
to be a quite good approximation.

The formulas (\ref{muh}) and (\ref{M}) are the good starting point for the
numerical calculations. From these formulas we see that in the limit $W/\mu
\ll 1$ the oscillating parts of magnetization and chemical potential are
connected by the simple relation 
\begin{equation*}
\tilde{M}(B)=\frac{\varepsilon _{F}}{B}\frac{2g}{\hbar \omega _{c}}(1+n_{R})%
\tilde{\mu}(B)
\end{equation*}
At zero warping this was obtained in \cite{we}.

The nonlinear equation (\ref{muh}) for $\tilde{\mu}(B)$\ can be solved
analytically only in some simple approximations. We shall do this to
illustrate the influence of the oscillations of the chemical potential on
the temperature and Dingle temperature dependence of the harmonic
amplitudes. So, we consider zero warping, zero spin-splitting and zero
temperature. Then the sum in the right-hand side of equation (\ref{muh}) can
be calculated and we get 
\begin{equation}
\frac{x}{2}=\frac{1}{(1+n_{R})}\arctan \left( \frac{\sin (y+x)}{\cos
(y+x)+e^{b}}\right)  \label{eqmu}
\end{equation}
where $x\equiv 2\pi \tilde{\mu}(B)/\hbar \omega _{c}$, $y\equiv 2\pi
\varepsilon _{F}/\hbar \omega _{c}$ and $b\equiv 2\pi \Gamma /\hbar \omega
_{c} $.

At very large electron reservoir $n_{R}=\infty $ , $x=0$ \ and we have the
limit of fixed chemical potential. In this case the magnetization is given
by \cite{Min} 
\begin{equation}
\tilde{M}(B)=\frac{2g\,\varepsilon _{F}}{\pi B}\arctan \left( \frac{\sin (y)%
}{e^{b}+\cos (y)}\right)  \label{Mbesc}
\end{equation}
The temperature dependence of the harmonic amplitudes is given by the L-K
formula: 
\begin{equation}
A_{l}(T)=\frac{2\pi ^{2}T\,l/\hbar \omega _{c}}{\sinh (2\pi ^{2}T\,l/\hbar
\omega _{c})}  \label{LK}
\end{equation}

It is possible to solve analytically the equation (\ref{eqmu}) also at $%
n_{R}=0$ and $n_{R}=1$. At zero electron reservoir $n_{R}=0$ the solution of
this equation is 
\begin{equation*}
\frac{x}{2}=\pi \frac{\tilde{\mu}(B)}{\hbar \omega _{c}}=\arctan \left( 
\frac{\sin (y)}{e^{b}-\cos (y)}\right)
\end{equation*}
It gives the oscillations of the chemical potential. The magnetization at
zero electron reservoir 
\begin{equation}
\tilde{M}(B)=\frac{2g\,\varepsilon _{F}}{\pi B}\arctan \left( \frac{\sin (y)%
}{e^{b}-\cos (y)}\right)  \label{Mzero}
\end{equation}
It coincides with (\ref{Mbesc}) after the phase shift $y\rightarrow y+\pi $
\ and the sign change $\tilde{M}\rightarrow -\tilde{M}$ . This means that
the harmonic damping law 
\begin{equation}
A_{l}\sim (1/l)\cdot \exp \left( -l\cdot b\right)  \label{Al}
\end{equation}
does not change, and only the sign of all even harmonics turns to inverse.
This symmetry between the cases of fixed chemical potential $\mu =const$ and
constant particle density $N=const$ is a feature of special exponential law
of harmonic damping. Any finite temperature and the density of electron
reservoir breaks this symmetry.

Let us consider now the intermediate case $n_{R}=1$. The equation (\ref{eqmu}%
) becomes 
\begin{equation}
\frac{\sin x}{\cos x}=\frac{\sin (y+x)}{\cos (y+x)+e^{b}}
\end{equation}
It gives 
\begin{equation*}
x=\arcsin \left( e^{-b}\sin y\right)
\end{equation*}
and the magnetization becomes 
\begin{equation}
\tilde{M}(y)=\frac{g\,\varepsilon _{F}}{\pi B}\arcsin \left( e^{-b}\sin
y\right)  \label{Mone}
\end{equation}
To say how the harmonic damping has changed we have to calculate the
amplitudes of several first harmonics of this expression.\bigskip\ The
amplitude of the first harmonic is 
\begin{equation*}
A_{1}(b)=\frac{1}{\pi }\int_{-\pi }^{\pi }\arcsin \left( e^{-b}\sin y\right)
\sin y\,dy
\end{equation*}
after integration by parts we get 
\begin{equation*}
A_{1}(b)=\frac{4}{\pi }\int_{0}^{\pi /2}\frac{\cos ^{2}y\,e^{-b}\,dy}{\sqrt{%
1-e^{-2b}\sin ^{2}y}}
\end{equation*}
This is the superposition of two elliptic integrals: 
\begin{equation}
A_{1}(b)=\frac{4}{\pi }\left[ e^{b}E(e^{-b})-2\sinh b\,K(e^{-b})\right]
\label{A1}
\end{equation}

At $b\gg 1$ the deviations of $A_{1}(b)$ from the L-K formula are small: 
\begin{equation*}
A_{1}(b)=e^{-b}+e^{-3b}/8+..
\end{equation*}
In the opposite limit $b\ll 1$ we get 
\begin{equation}
A_{1}(b)=\frac{4}{\pi }\left\{ 1-b\left( \ln \frac{4}{\sqrt{2b}}-\frac{1}{2}%
\right) +O(b^{2})\right\}  \label{A10}
\end{equation}
This is substantially different from the L-K dependence $A_{1}(b)=\exp
(-b)\approx 1-b$. For example, the value $A_{1}(0)$ is $4/\pi $ times larger
than the L-K prediction.

The amplitudes of the next harmonics reveal the stronger deviation from the
L-K formula (\ref{Al}). All even harmonics disappear since the expression (%
\ref{Mone} ) has the symmetries $\tilde{M}(\pi -y)=\tilde{M}(y)$ and $\tilde{%
M}(-y)=\tilde{M}(y).$

The amplitude of third harmonic can also be calculated. At $b\gg 1$, $%
e^{-b}\ll 1,$%
\begin{equation*}
A_{3}(b)=-e^{-3b}/12+O(e^{-5b})
\end{equation*}
This result is in contrast to the cases $n_{R}=0$ \ or $n_{R}=\infty $ where
we had $A_{3}(b)=e^{-3b}/3$ . It is not surprising since in the symmetric
case $n_{R}=1$ the oscillations should be much smoother and more sinusoidal.
Hence one should have an increase of first harmonic and a decrease of higher
harmonics. At $b=0$%
\begin{equation}
A_{3}(0)=\frac{4\,}{3\pi }\int_{0}^{\pi /2}\frac{\cos 3y\,\cos y\,dy}{\cos y}%
=-\frac{4}{9\pi }  \label{A30}
\end{equation}
which is $\sim 2.35$ times less than the Lifshitz-Kosevich prediction $%
A_{3}(0)=1/3$ and has the opposite sign.\ So, in the case $n_{R}=1$\ the
first harmonic is increased while the others are strongly decreased in
amplitude compared to the cases of zero and infinite electron reservoir. The
deviation from the L-K formula reduces as the warping of FS increases. The
above analysis shows also that the harmonic ratios at low temperature and
Dingle temperature can give a quantitative estimate of the electron
reservoir density which is much more precise than just a note about the
slope of the magnetization curve.

\FRAME{ftbpFU}{3.5354in}{2.7916in}{0pt}{\Qcb{{}Temperature dependence of
harmonic amplitudes. The solid lines are the numerical results (at $n_{R}=1$%
, $m^{\ast }=2m_{0}$, $T_{D}=0.2K$ and $W=1K$; see text) and the dashed
lines represent the Lifshitz-Kosevich prediction at the same parameters.
Their strong deviations are clearly seen, especially for higher harmonics.}}{%
\Qlb{Hal}}{hal.eps}{\special{language "Scientific Word";type
"GRAPHIC";maintain-aspect-ratio TRUE;display "USEDEF";valid_file "F";width
3.5354in;height 2.7916in;depth 0pt;original-width 4.1805in;original-height
3.1479in;cropleft "0";croptop "1.0467";cropright "1";cropbottom "0";filename
'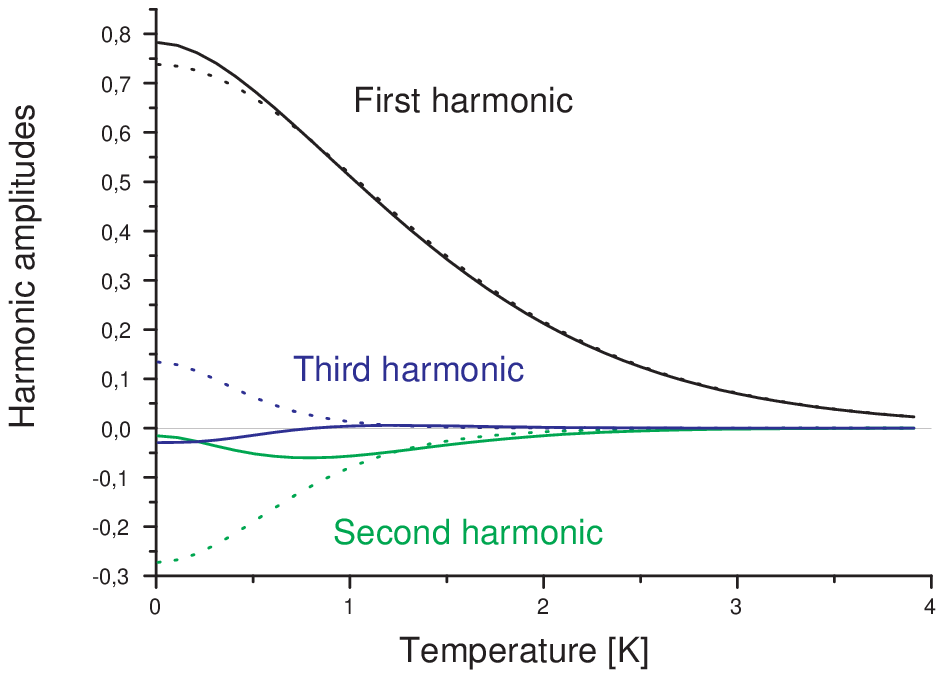';file-properties "XNPEU";}}

To include correct temperature dependence, warping and spin-splitting \ and
consider arbitrary reservoir density one can perform the numerical
calculations, based on the solution of the equation (\ref{muh}) for the
chemical potential and substitution of this solution into the formula (\ref
{M}) for the magnetization. This can be done at arbitrary parameters,
available on the experiment. The temperature dependence of the first three
harmonic amplitudes is given in fig.1 for the following set of parameters,
close to the real experiments on $\alpha $-$(BEDT-TTF)_{2}KHg(SCN)_{4}$: the
reservoir density $n_{R}=1$, the dHvA frequency $F=700$ Tesla, the effective
mass $m^{\ast }=2m_{0}$, the Dingle temperature $T_{D}=0.2K$ and the warping 
$W=1K$. We see the substantial deviation from the Lifshitz-Kosevich
dependence. The obtained amplitude of the first harmonic at $T\rightarrow 0$
is about $1.1$ times larger than the Lifshitz-Kosevich prediction. If we put
also $T_{D}\rightarrow 0$ and $W\rightarrow 0$ their ratio becomes $4/\pi
=1.27$ in agreement with the analytical result (\ref{A10}). The second
harmonic amplitude is close to zero at $T=0$. The amplitude of the third
harmonic changes the sign at $T\approx 0.8K$ and deviates very strongly from
the L-K formula. It is damped much stronger than the L-K predictions. At $%
T=0 $ and $W=0$ it also coincides with the prediction (\ref{A30}).

To conclude, it was shown both analytically and numerically that the
oscillations of the chemical potential are essential for the temperature
dependence of harmonic amplitudes of dHvA oscillations in
quasi-two-dimensional compounds. The accurate determination of the effective
electron mass from the dHvA effect should take this effect into account.
This can be done by the simple numerical calculation based on the formulas (%
\ref{muh}) and (\ref{M}). The oscillations of the chemical potential depend
on the reservoir density of states according to the formula (\ref{muh}).
This fact may be used for the estimate of the reservoir density of states in
organic metals.

\bigskip The author thanks A.M. Dyugaev and M.V. Kartsovnik for useful
discussions. The work was supported by RFBR grant N 00-02-17729a. 

\end{document}